\documentstyle[12pt]{article}

\begin{document}

\title{Quantum-defect Electron States Near a Helium Surface }
\author{B.A. Lysov, O.A. Lysova, O.F. Dorofeyev \\
Physical Faculty, Moscow State University,\\
119899, Moscow, Russia}
\maketitle

\begin{abstract}
Isolated electrons resting near a helium surface have a spectrum close to
that of a quantum-defect atom. A precisely solvable model with Ryd\-berg
spectrum is sugguested and discussed.

PACS: 03.65.Ge, 73.20-r

E-mail:dorofeev@vega.phys.msu.su
\end{abstract}

\qquad Recently M.M. Nieto from Los Alamos National Laboratory has noticed
that observed spectrum of the electrons resting near a helium surface is
close to electrons resting near a helium surface is close to that of
quantum-defect Rydberg atom

\begin{equation}
E=\;-\,\;\frac{{\cal E}_{0}}{\left( n-\delta \right)^{2}}\\;, 
\label{1}
\end{equation}
where ${\cal E}_{0}=158.4\;GHz$ and $\delta =0.0237$ \cite{1}. He has suggested a
new mathematical model to describe this phenomena. The remarkable Nieto's
model permits to solve the problem in analytic closed form, but
unfortunately it is somewhat dubious from purely physical point of view. In
these notes we call attention to another precisely solvable model with
Rydberg spectrum. Let us write the Schroedinger equation for the problem
under consideration 
\begin{equation}
\left( -\frac{\hbar ^{2}}{2m}\;\frac{d^{2}}{dx^{2}}\;-\;\frac{Ze^{2}}{x}%
\right) \;\Phi \left( x\right) =\;E\;\Phi \left( x\right) ,\qquad Z=\frac{%
1-\varepsilon }{4\left( 1+\varepsilon \right)}.    \label{2}
\end{equation}
Here $\varepsilon $=1.05723 is dielectric constant of helium. Our main idea
is that one has to solve equation (2) on whole coordinate $x-$ axis but
not only on its positive part. A solution of equation (2) that is continuous
everywhere on coordinate $x-$ axis and has the right behavior both at + and
- infinity can be written in the form 
\[
\Phi \left( x\right) =N\;\left\{ 
\begin{array}{cc}
\Gamma \left( s-1\right) W_{s,\;1/2}\left( z\right) ,\;\; & \;x\geq 0, \\ 
\Gamma \left( s+1\right) W_{-\;s,\;1/2}\left( z\right) , & \;x\leq 0.
\end{array}
\right. 
\]
where $N$ is the normalization constant, $W_{s,\;1/2}\left( z\right) $ is
Whittaker confluent hypergeometric function, 
\[
z=\frac{x}{sx_{0}},\quad s>0,\quad x_{0}=\frac{\hbar ^{2}}{2mZe^{2}},
\]
and the index $s$ is related with the eigenvalue $E$ by the equality 
\[
E=-\;\frac{{\cal E}_{0}}{s^{2}},\quad {\cal E}_{0}=\frac{mZ^{2}e^{4}}{2\hbar
^{2}}.
\]
To determine the eigenvalues we impose on the wave functions the boundary
condition at singular point $x=0$ 
\begin{equation}
\left( \frac{d\,\Phi }{d\,x}\right) _{x\rightarrow \;+\;0}-\left( \frac{%
d\,\Phi }{d\,x}\right) _{x\rightarrow \;-\;0}=\frac{\lambda }{x_{0}}\Phi
\left( 0\right) ,  
\end{equation}
here $\lambda $ is an arbitrary dimensionless constant.

If we now make \ use of well know identity for polygamma functions 
\[
\Psi \left( z+1\right) -\Psi \left( z-1\right) =\frac{1}{z}-\pi \cot \left(
\pi z\right) 
\]
and take into account the asymptotic equalities [2]\ 
\[
\frac{x_{0}}{N}\;\Phi ^{\prime }\left( z\right) =-2C+\frac{1}{2s}-\ln
\;\left| z\right| -\psi \left( 1-s\right) +O\left( z\ln z\right) ,\quad z>0, 
\]
\[
\frac{x_{0}}{N}\;\Phi ^{\prime }\left( z\right) =-2C-\frac{1}{2s}-\ln
\;\left| z\right| -\psi \left( 1+s\right) +O\left( z\ln z\right) ,\quad z<0, 
\]
where $C$ is Euler's constant, which may be easily established with the aid
of \cite{2} we shell easily find 
\[
\cot \left( \pi s\right) =-\;\frac{\lambda }{\pi }. 
\]
Putting now 
\[
\lambda =\pi \cot \,\left( \pi \delta \right) ,\qquad 0<\delta <1, 
\]
where $\delta $ is a \ new arbitrary constant, we get 
\[
s=n-\delta ,\qquad n=1,2,... 
\]
and for he eigenvalues we have the same formula (1) with the slightly
different value ${\cal E}_{0}=159.123\;GHz$ \ There is an also easy task to
calculate normalization constant $C$ and mean value $\left\langle
x\right\rangle .$ The results are 
\[
N=\left[ \sqrt{2x_{0}s}\,\Gamma \left( 1-s\right) \,\Gamma \left( 1+s\right)
\right] ^{-1}, 
\]
\[
\left\langle x\right\rangle _{n}=x_{0}\,\left[ 3n^{2}-\delta \left(
2n-\delta \right) \right] . 
\]
Thus, we have an exact analytic solution to the problem. The boundary
condition (3) may be hidden into Schroedinger equation itself if we
introduce additional short range potential of the form 
\begin{equation}
V_{short\;range}\left( x\right) =\lambda \,Z\,e^{2\,}\delta \left( x\right) .
  \label{4}
\end{equation}
We think that it ``realistically'' models the positive work function at the
surface of about 1eV \cite{3}. The transmission coefficient for potential (4) is
given by 
\begin{equation}
T=\left[ 1+\lambda ^{2}{\cal E}_{0}/E\right] ^{-1}.    \label{5}
\end{equation}

Let us denote 
\[
\stackrel{\_}{E}\;=\lambda ^{2\,}{\cal E}_{0}. 
\]
If the electron energy exceed $\stackrel{\_}{E}$ the probability to
penetrate the liquid - gas interface according to (5) will be more than $0.5$. For
above mentioned values of \ $\delta $ and ${\cal E}_{0}$ we have $\stackrel{%
\_}{E}=1.1687\;eV$.

\bigskip

\section*{Acknowledgments}

We are grateful to V.Ch. Zhukovskii and A.V. Borisov for helpful discussions.


\bigskip

\bigskip

\begin{thebibliography}{5}

\bibitem{1}M.M. Nieto, quant-ph/9908059.
\bibitem{2}A.P. Prudnikov, Yu.A. Brychkov, O.I. Marichev. Integraly i ryady.
\ \ Dopolnitel'nye glavy. \ Moscow. Nauka. 1986.
\bibitem{3}D.K. Lambert and P.L. Richards, Phys. Rev. Lett. {\bf 44}. 1427\ (1980);\ \
Phys. Rev. {\bf B23}. 3282 (1981).

\end{thebibliography}
\end{document}